\documentclass[twocolumn,preprintnumbers,amsmath,amssymb]{revtex4}
\usepackage{graphicx}% Include figure files
\usepackage{dcolumn}% Align table columns on decimal point
\usepackage{bm}% bold math

\begin{document}
\title{Entanglement in a hardcore-boson Hubbard model}

\author{Xiang Hao}

\author{Shiqun Zhu}
\altaffiliation{Corresponding author} \email{szhu@suda.edu.cn}

\affiliation{School of Physical Science and Technology, Suzhou
University, Suzhou, Jiangsu 215006, People's Republic of China}

\begin{abstract}

The entanglement in a Hubbard chain of hardcore bosons is
investigated. The analytic expression of the global entanglement in
ground state is derived. The divergence of the derivative of the
global entanglement shows the quantum criticality of the ground
state. For the thermal equilibrium state, the bipartite and the
multipartite entanglement are evaluated. The entanglement decreases
to zero at a certain temperature. The thermal entanglement is
rapidly decreasing with the increase of the number of sites in the
lattice. The bipartite thermal entanglement approaches a constant
value at a certain number of sites while the multipartite
entanglement eventually vanishes.


\end{abstract}

\maketitle

\section{Introduction}

Quantum entanglement is an important property which plays an
essential role in the quantum information processing
\cite{Divin00,Cerf02,Brukner02}. There are different types of
entanglement, such as bipartite, multipartite, block
\cite{Latorre03}, and localizable entanglement, etc \cite{Popp05}.
The relative entropy of entanglement \cite{Bennett96} and the
entanglement of formation \cite{Wootters98} can be used for qubits.
Arbitrary bipartite entanglement can be assessed by the "negativity"
\cite{Vidal02}. In recent years, the entanglement in Heisenberg
models of finite systems of spins has been investigated
\cite{Arnesen01,Kamta02,Santos03,Subra04,Gu04,Wang02,Jaeger03,Brennen04,Endrejat05}.
The anisotropy effect \cite{Kamta02,Santos03}, multi-dimensional
lattices \cite{Subra04}, long-range interactions \cite{Gu04},
multipartite entanglement
\cite{Wang02,Jaeger03,Brennen04,Endrejat05} have all been studied in
Heisenberg models. The entanglement in solids can also be witnessed
by the magnetization \cite{Wang022} and the thermal energy
\cite{Dowling04,Toth05,Lunkes05}. In solids, there is a
characteristic temperature below which a thermal entangled state can
be obtained. The effects of quantum entanglement have been detected
in the experimental situation \cite{Ghosh03}. Some attention has
been drawn to the entanglement in the models of infinite spin
systems \cite{Amico04,Korepin04}. Owing to the quantum nonlocal
correlations, the connection of the entanglement and quantum
criticality in spin systems has been discussed
\cite{Osborne02,Eisenberg04,Fan04,Briegel05,Wu04,Somma04,Wei05,Larsson05,Brandes04,Lambert05}.
The quantum criticality can be shown by the entanglement of the
ground state \cite{Latorre03,Wei05}. Next to Heisenberg models,
boson Hubbard models have been extensively used to study the
metal-insulator transition \cite{Hebert01}. The Hubbard chain of
hardcore bosons is one of the simplest models which can embody such
quantum criticality. It is of interest to investigate the
entanglement in the hardcore boson Hubbard chain.

In this paper, the entanglement of a hardcore boson Hubbard chain is
studied for the ground and thermal states. In Sec. II, the global
entanglement measure of the ground state is analytically expressed
by the magnetization. The critical properties of the transition
between the Mott-insulating phase and the superfluid phase are shown
by the derivative of the global entanglement at ground state. In
Sec. III, the "negativity" of any two sites in the chain is derived.
The multipartite entanglement at finite temperature is calculated.
The effects of the number of lattice sites, the temperature, and the
chemical potential are investigated. A discussion concludes the
paper.

\section{Global entanglement of ground state}

Recently, the quantum entanglement of strongly correlated spin
systems has been extensively studied. One of the interesting focuses
was the relation of quantum phase transitions and entanglement in
the ground state. As is well known, the phenomenon of quantum phase
transitions describes the global properties of the ground state.
Therefore, the measurement of global quantum entanglement is
relevant to the investigation of the quantum phase transition. For a
typical case, the spin version of the one-dimensional boson Hubbard
is expressed by
\begin{equation}
\label{eq:1} H=-w \sum_{\langle ij
\rangle}(a_{i}^{\dag}a_{j}+a_{j}^{\dag}a_{i})-\mu \sum_{i}
n_{ai}+U\sum_{i} n_{ai}(n_{ai}-1)
\end{equation}
where $a_{i}^{\dag}$ and $a_{i}$ are the creation and annihilation
operator for bosons on the site $i$ of the lattice,
$n_{ai}=a_{i}^{\dag}a_{i}$ is the number operator. The parameter
$w>0$ allows hopping of bosons from one site to another, $\mu$
represents the chemical potential of the bosons, and $U$ denotes the
possible repulsive interaction among bosons on each site. For
simplicity, only the nearest-neighbor pairs $\langle ij \rangle$ are
considered. The off-site and long-range repulsive interactions are
neglected. If the repulsion is very strong, when $U\rightarrow
\infty$, there is only one boson at each site. The model is reduced
to the simplest one of a hardcore boson Hubbard chain, which can
also be written as a magnetic model of $S=\frac 12$ spin with
pairwise interaction. The relation
$\sigma_i^{z}=1-2a_{i}^{\dag}a_{i}$ is satisfied. In this following,
the hardcore boson Hubbard chain is studied. It is known that this
model is equivalent to the spin-$1/2$ Heisenberg XX chain with the
ferromagnetic interaction $-w/2$ and the external magnetic field
$\mu$. Through the Jordan-Wigner and Fourier transformations, the
Hamiltonian $H$ can be exactly expressed by
\begin{equation}
\label{eq:2} H=-\sum_k \epsilon_k c_k^{\dag}c_k
\end{equation}
where $\epsilon_k=-2w\cos\frac {2\pi k}{L}-\mu$, $L$ is the number
of sites on the lattice, and $c_k^{\dag},c_k$ are the
Fourier-transformed fermionic operators.

To show the relation of entanglement and quantum criticality, the
global entanglement $E$ of the ground state can be introduced by
\cite{Meyer02, Brennen03},
\begin{equation}
\label{eq:3} E=2[1-\frac 1L\sum_{j=1}^Ltr(\rho_j^2)]
\end{equation}where $tr(\rho_j^2)$ is the trace of the reduced density matrix
$\rho_i$ on the $i$th site of the ground state. Here, the
eigenstates of $\sigma_i^z$ are assumed to be
$\{|0\rangle_i,|1\rangle_i\}$. The reduced density matrix is given
by $\rho_i=\frac {I+M \sigma_i^z}2$ where $M$=$\frac
1L\sum_{i}^{L}tr(\rho_i \sigma_i^z)$. Therefore, the global
entanglement $E$ of the ground state can be just obtained by
\begin{equation}
\label{eq:4} E=1-M^2
\end{equation}
For the values of $|\mu|\leq 2w$, there is partial occupation of
sites at the ground state and $M=\frac 2{\pi}\cos^{-1}\frac
{\mu}{2w}-1$. When $|\mu|\geq 2w$, $|M|=1$. The global entanglement
of the ground state is plotted as functions of the hopping
coefficient $w$ and the potential $\mu$ in Fig. 1(a) when the number
of sites is $L=10^{4}$ . It is shown that the global entanglement
$E$ exists at the ground state if the potential satisfies $|\mu|\leq
2w$. The values of $E$ are decreasing in $|\mu|$ and then drop to
zero when $|\mu|=2w$. When $\mu\rightarrow 0$, it is the maximally
entangled ground state which is exactly the
Greenberger-Horne-Zeilinger state of the form $|\psi\rangle_g=\frac
1{\sqrt2}(|010\cdots1\rangle+|101\cdots0\rangle)$. If the values
$\mu \leq -2w$ or $\mu \geq 2w$, the ground state is an unentangled
pure state $|00\cdots0\rangle$ or $|11\cdots1\rangle$. According to
\cite{Hebert01}, there are two different kinds of phases namely the
Mott-insulating phase and the superfluid one. It is found that the
transitions between them occur under the condition $\mu=\pm 2w$. The
global entanglement always exists in the superfluid phase while
there is no entanglement in the Mott-insulating phase. To clearly
demonstrate the phenomenon of a quantum phase transition in the
ground state, the derivative of the global entanglement
$E^{'}_{\mu}$ is obtained
\begin{equation}
\label{eq:5} E^{'}_{\mu}=\frac {\partial E}{\partial \mu}=\frac
4{\pi^2\sqrt{4w^2-\mu^2}}[2\cos^{-1}\frac {\mu}{2w}-\pi]  ,\quad
|\mu|\leq 2w.
\end{equation}
When the potential $\mu\rightarrow 2w_{-}$, $\frac {\partial
E}{\partial \mu}\rightarrow -\infty$ which reveals the divergence of
$\frac {\partial E}{\partial \mu}$. The quantum criticality is
depicted in Fig. 1(b) when the hopping coefficient is chosen to be
$w=1$. If $\mu\rightarrow 2w_{+}$, the derivative $\frac {\partial
E}{\partial \mu}=0$ for the global entanglement $E=0$. It is found
that the quantum criticality in the hardcore boson Hubbard chain can
be shown by the global entanglement at the ground state.

\section{Entanglement at finite temperature}

The thermal equilibrium state is $\rho(T)=e^{-H/kT}/Z$, where $Z$ is
the partition function at finite temperature $T$ and $k$ is the
Boltzmann constant. For convenience, both the Boltzmann constant $k$
and the Planck constant $\hbar$ are assumed to be one. Because the
Hubbard chain of hardcore bosons is equivalent to the spin-$1/2$
Heisenberg XX chain, the reduced density matrix $\rho_{ij}$ on any
two sites $i$ and $j$ can be expressed by the correlation function
$K_{ij}^{\alpha \alpha}=tr(\rho \sigma_i^{\alpha}
\sigma_j^{\alpha}), (\alpha=x,y,z)$. In the Hilbert space of
$\{|00\rangle_{ij},|01\rangle_{ij},|10\rangle_{ij},|11\rangle_{ij}\}$,
the expression for $\rho_{ij}$ can be obtained
\begin{equation}
\label{eq:6} \rho_{ij}=\left(\begin{array}{cccc}
                              u & 0 & 0 & 0\\
                              0 & w & t & 0\\
                              0 & t & w & 0\\
                              0 & 0 & 0 & v
                              \end{array}\right)
\end{equation}
where $u=\frac 14(K_{ij}^{zz}+2M+1)$, $v=\frac
14(K_{ij}^{zz}-2M+1)$, $w=\frac 14(1-K_{ij}^{zz})$, and $t=\frac 12
K_{ij}^{xx}$. The analytical calculations of correlation functions
$K_{ij}^{\alpha \alpha}$ and the magnetization $M$ are
straightforward given. For the number of sites $L$, the
magnetization is given by $M=-\frac
1L\sum_{q=1}^{L}\tanh(\epsilon_q/2T)$. The two-site correlations can
be by
\begin{eqnarray}
\label{eq:7} K_{ij}^{xx}& = &K_{ij}^{yy}=\left|\begin{array}{cccc}
                               G_1 & G_0 & \cdots & G_{-r+2}\\
                               G_2 & G_1 & \cdots & G_{-r+3}\\
                               \vdots & \vdots & \ddots & \vdots\\
                               G_r & G_{r-1} & \cdots & G_1
                               \end{array}\right|,\nonumber \\
             K_{ij}^{zz}& = & 4M^2-G_rG_{-r}
\end{eqnarray}
where the parameter $r$=$|j-i|$ is the separation distance between
two sites, and the item $G_r$=$G_{-r}$=$\frac
1L\sum_{q=1}^{L}\cos(2\pi qr/L)\tanh(\epsilon_q/2T)$.

The thermal entanglement in the chain can be investigated by the
negativity $N$. Based on the separability principle, the
negativity $N$ can be used to quantify the bipartite entanglement
between two sites \cite{Vidal02}. The negativity $N$ is introduced
by
\begin{equation}
\label{eq:8} N(\rho)=2|\sum_i\lambda_i|
\end{equation}
where $\lambda_i$ is the $i$th negative eigenvalue of $\rho^{T}$
which is the partial transpose of the mixed state $\rho$. From the
separability of quantum states, the partial transpose matrix
$\rho^{T}$ has nonnegative eigenvalues if the states are unentangled
and the value $N(\rho)=0$. In the Hubbard model of hardcore bosons,
the negativity of the two-site entangled state is given by
\begin{equation}
\label{eq:9} N(\rho_{ij})=|u+v-\sqrt{(u-v)^2+4t^2}|
\end{equation}
Thus, the entanglement on any two sites can be calculated
numerically through Eqs. (\ref{eq:7})-(\ref{eq:9}). The negativity
$N$ is plotted as a function of site number $L$ in Fig. 2(a) when
the temperature is $T=0.5$, the hopping coefficient $w=1$, and the
potential $\mu=0.2$. It is found that the pairwise entanglement
always exists whatever the number $L$ of sites is. There is no
thermal entanglement on two sites when the separation distance is
$r>2$. The value of $N$ is rapidly decreasing in $L$, and then
reaches a constant value at $L=16$. This result illustrates that the
thermal entanglement can be detected in real solids of a very large
number of particles.

The multipartite entanglement $E_L$ for the thermal states of
systems with an even number of sites $L$ can be introduced
\cite{Brennen04}
\begin{equation}
\label{eq:10} E_L(\rho)=\max\{0,
\nu_{0}-\sum_{j=1}^{2^L-1}\nu_{j}\}
\end{equation}
where $\{\nu_{j}\}_{j=0}^{2^n-1}$ is the spectrum of the operator
$\sqrt{\rho U \rho U^{-1}}$ in decreasing order, $U$ is an
anti-unitary time reversal operator and can be written as
$U=[\prod_{j=1}^{L}(-i\sigma_{j}^{y})]\tau$, $\tau$ is the complex
conjugate operator. The multipartite entanglement $E_L$ is plotted
in Fig. 2(b). In Fig. 2(b), the value of $E_L$ is decreased with the
increase of the number of sites $L$. The multipartite entanglement
$E_L$ vanishes at $L=10$. However, the bipartite entanglement still
exists in this case.

It is also very interesting to study the effects of the temperature
$T$ and the chemical potential $\mu$ on the thermal entanglement. By
the analytical expression for the bipartite entanglement in Eq.
(\ref{eq:9}), the negativity $N$ is plotted as a function of $T$ and
$\mu$ in Fig. 3(a) for $w=1$ and $L=10^{4}$. When the temperature
$T$ and the chemical potential $|\mu|$ are increased, the negativity
$N$ decreases. It is found that the values of the negativity $N$ are
symmetric about the chemical potential $\mu$. When $\mu\rightarrow
0$, the value of $N$ is maximal. For a definite chemical potential,
the bipartite entanglement disappears at a certain temperature
$T_c$. It is clear that the values of $T_c$ can be increased by
decreasing the chemical potential $|\mu|$. It is seen that the
entanglement can be detected at low temperatures in solids. The
multipartite entanglement $E_L$ for the thermal state of $L=6$ is
plotted in Fig. 3(b) when $w=1$. It is shown that the values of
$E_L$ are also symmetric about the chemical potential $\mu$ and
decreased with the increase of $|\mu|$. The ground state for
$|\mu|<0.5$ is just the maximally entangled GHZ state $\frac
1{\sqrt2}(|010\cdots1\rangle+|101\cdots0\rangle)$. The values of
$E_L$ are declined with the temperatures $T$ and vanishes at about
$T=0.6$.

\section{Discussion}

The entanglement in a hardcore boson Hubbard chain at ground and
thermal equilibrium states is investigated. The global entanglement
at ground state is analytically expressed by the magnetization. When
the potential $\mu\rightarrow 0$, the maximally entangled
Greenberger-Horne-Zeilinger state can be obtained. The quantum
criticality is revealed by the divergence of the derivative of the
global entanglement at ground state. In the parameter plane of
Mott-insulator and superfluidity phases, it is found that the
entanglement exists in the SF phase while there is no entanglement
in the MI phase. The bipartite entanglement between any two sites is
deduced by the negativity. For a very large number of sites $L$, the
pairwise entanglement can always exist. When the number $L$
increases, the negativity decreases rapidly and then reaches a
constant value at a certain number of site. While the multipartite
entanglement will decrease to zero with increasing $L$. The thermal
entanglement vanishes at a certain temperature and is decreased with
the increase of the potential $\mu$. It is shown that the
entanglement can be detected at low temperature in real solids of a
large number of sites.

\vskip 0.5cm

{\large \bf Acknowledgement}

It is a pleasure to thank Yinsheng Ling, Jianxing Fang, and Qing
Jiang for their many fruitful discussions about the topic. The
financial support from the Special Research Fund for the Doctoral
Program of Higher Education (Grant No. 20050285002) is gratefully
acknowledged.

\newpage

{\large \bf Figure Captions}

\vskip 0.5cm

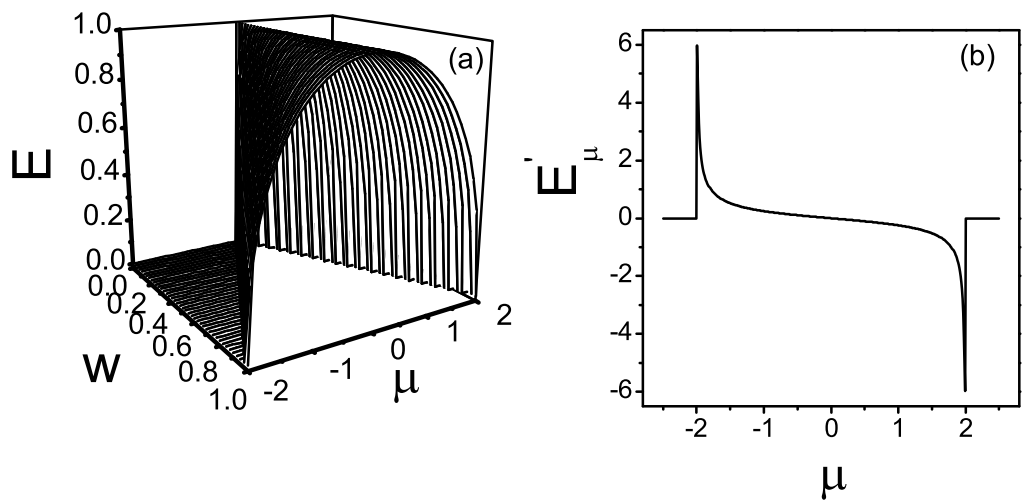
{\bf Fig. 1}

(a) The global entanglement $E$ of the ground state is plotted as a
function of the hopping coefficient $w$ and the potential $\mu$. The
number of sites is $L=10^4$; (b)The derivative $\frac {\partial
E}{\partial \mu}$ is plotted to show the quantum criticality of the
ground state.

\vskip 0.5cm

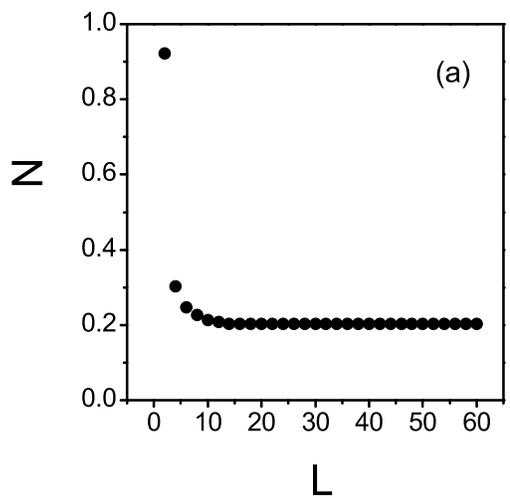
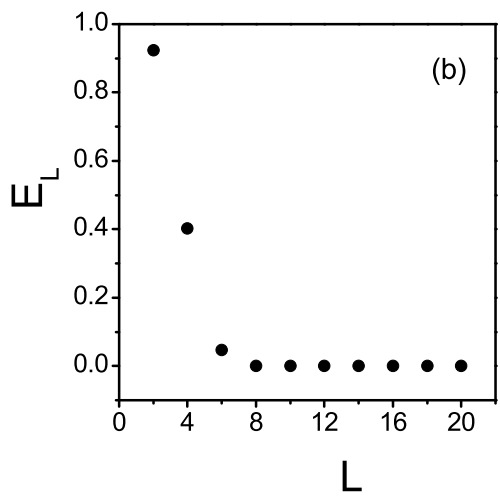
{\bf Fig. 2}

The thermal entanglement is plotted as a function of the number of
sites when $w=1, \mu=0.2$, and $T=0.5$. (a) The pairwise
entanglement of the negativity $N$; (b) The multipartite
entanglement $E_L$.

\vskip 0.5cm

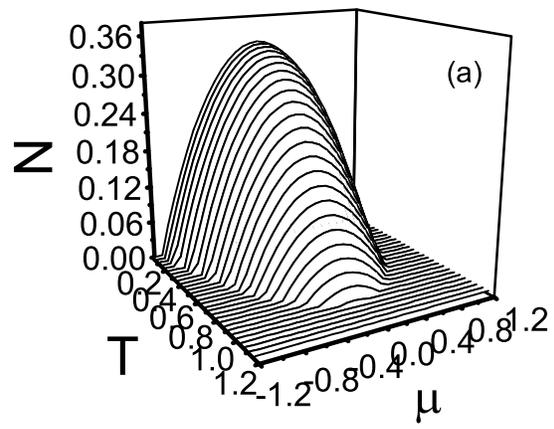
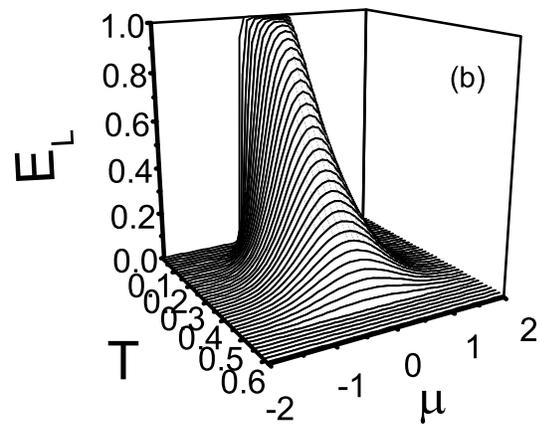
{\bf Fig. 3}

The thermal entanglement is plotted as a function of the potential
$\mu$ and the temperature $T$ when the hopping coefficient is $w=1$.
(a) The pairwise entanglement for $L=10^{4}$; (b) The multipartite
entanglement $E_L$ for $L=6$.

\end{document}